\documentclass{article}

\usepackage{microtype}
\usepackage{graphicx}
\usepackage{subcaption}
\usepackage{booktabs} 

\usepackage{overpic}
\usepackage{xcolor} 

\usepackage{hyperref}

\newcommand{\icmltitle}{\mlforastrotitle}
\newcommand{\icmltitlerunning}{\mlforastrotitlerunning} 
\newcommand{\icmlauthor}{\mlforastroauthor}
\newcommand{\icmlaffiliation}{\mlforastroaffiliation} 
\newcommand{\icmlcorrespondingauthor}{\mlforastrocorrespondingauthor}
\newcommand{\icmlkeywords}{\mlforastrokeywords}

\newenvironment{icmlauthorlist}
  {\begin{mlforastroauthorlist}}
  {\end{mlforastroauthorlist}}
\usepackage[accepted]{ml4astro2025}

\usepackage{amsmath}
\usepackage{amssymb}
\usepackage{mathtools}
\usepackage{amsthm}

\usepackage[capitalize,noabbrev]{cleveref}

\theoremstyle{plain}

\theoremstyle{definition}

\theoremstyle{remark}

\usepackage[textsize=tiny]{todonotes}

\icmltitlerunning{Bridging Simulators with Conditional Optimal Transport}

\begin{document}
\newcommand{\ju}[1]{\textcolor{cyan}{{ju: #1}}}

\twocolumn[
\icmltitle{Bridging Simulators with Conditional Optimal Transport}




\begin{icmlauthorlist}
\icmlauthor{Justine Zeghal}{1,2,3}
\icmlauthor{Benjamin Remy}{4}
\icmlauthor{Yashar Hezaveh}{1,2,3,7,8}
\icmlauthor{François Lanusse}{5,6}
\icmlauthor{Laurence Perreault-Levasseur}{1,2,3,7,8,9}
\end{icmlauthorlist}

\icmlaffiliation{1}{Department of Physics, Université de Montréal, Montréal, Canada}
\icmlaffiliation{2}{Mila – Quebec Artificial Intelligence Institute, Montréal, Canada}
\icmlaffiliation{3}{Ciela – Montreal Institute for Astrophysical Data Analysis and Machine Learning, Montréal, Canada}
\icmlaffiliation{4}{Department of Astrophysical Sciences, Princeton University, Princeton, NJ, USA}
\icmlaffiliation{5}{Université Paris-Saclay, Université Paris Cité, CEA, CNRS, AIM, Gif-sur-Yvette, France}
\icmlaffiliation{6}{Flatiron Institute, New York, NY, USA}
\icmlaffiliation{7}{Center for Computational Astrophysics, Flatiron Institute, New York, USA}
\icmlaffiliation{8}{Trottier Space Institute, McGill University, Montreal, Canada}
\icmlaffiliation{9}{Perimeter Institute for Theoretical Physics, Waterloo, Canada}

\icmlcorrespondingauthor{Justine Zeghal}{justine.zeghal@umontreal.ca}

\icmlkeywords{Machine Learning, ICML}

\vskip 0.3in
]



\printAffiliationsAndNotice{} 

\begin{abstract}
We propose a new field-level emulator that bridges two simulators using unpaired simulation datasets. Our method leverages a flow-based approach to learn the likelihood transport from one simulator to the other. Since multiple transport maps exist, we employ Conditional Optimal Transport Flow Matching (COT-FM) to ensure that the transformation minimally distorts the underlying structure of the data. We demonstrate the effectiveness of this approach by bridging weak lensing simulators: a Lagrangian Perturbation Theory (LPT) to a N-body Particle-Mesh (PM).  
We demonstrate that our emulator captures the full correction between the simulators by showing that it enables full-field inference to accurately recover the true posterior, validating its accuracy beyond traditional summary statistics. 

\end{abstract}

\section{Introduction}
In recent years, there has been a growing shift from traditional analytic inference methods to Simulation-Based Inference (SBI) \citep[e.g.][]{fluri2022full,porqueres2023fieldlevelinferencecosmicshear, jeffrey2024darkenergysurveyyear}. This shift is driven by SBI's ability to provide more precise constraints on cosmological parameters, particularly when analyzing data at the pixel level. However, it comes with the downside of relying solely on simulation accuracy. Any mismatches between the simulated and true underlying physical processes can result in biased parameter estimates \citep[e.g.][]{filipp2024robustnessneuralratioposterior, bayer2025fieldlevelcomparisonrobustnessanalysis}.
Although significant effort has been dedicated to improving the realism of simulations, a major challenge remains: high-fidelity simulations are computationally expensive, limiting both the number of samples and the size of each simulation. This constraint makes full-field inference difficult to apply without compromising simulation quality.

Emulators have emerged as a way to accelerate the generation of high-fidelity simulations. Neural network-based emulators are trained on simulations to either correct
fast approximate simulations
\citep[e.g.][]{Dai_2018, Kodi_Ramanah_2020, lanzieri2022hybridphysicalneuralodesfast, payot2023learningeffectiveevolutionequation, Bartlett_2025} 
or emulate the full simulations evolution
\citep[e.g.][]{Lucie_Smith_2018, He_2019, Jamieson_2023}. As learning a small correction between Low-Fidelity (LF) and High-Fidelity (HF) simulations should require fewer simulations, we focus on the formal approach. Specifically, we propose an emulator based on flow models \citep{lipman2023flowmatchinggenerativemodeling, pooladian2023multisampleflowmatchingstraightening, 
albergo2023stochasticinterpolantsunifyingframework, tong2024improvinggeneralizingflowbasedgenerative, lipman2024flowmatchingguidecode} that learn mappings between two probability distributions: the HF distribution and LF distribution. Since many mappings can transform one distribution into another, additional structure is needed to ensure that the correction minimally distorts the data manifold and maintains the correct correspondence between data samples and parameters.
To achieve this, we leverage Optimal Transport (OT) theory, which seeks to find minimal-effort mappings between distributions according to a chosen cost function. Specifically, we use Conditional Optimal Transport Flow
Matching (COT-FM)  \citep{kerrigan2024dynamicconditionaloptimaltransport}, an approach that aligns conditional distributions (i.e., likelihoods rather than marginals), ensuring that the mapping between input parameters and observables is preserved. Meanwhile, Flow Matching (FM) provides a framework for learning deterministic mappings by minimizing a regression loss on velocity fields, remaining efficient even in high dimensions. A key distinction of our approach compared to standard field-level emulators is that it does not require paired data (e.g., simulations with shared initial conditions), a flexibility enabled by our OT-based framework \citep{tong2024improvinggeneralizingflowbasedgenerative}.

We apply our COT-FM emulator to learn the transformation that maps second-order Lagrangian Perturbation Theory (LPT) convergence maps to their N-body Particle-Mesh (PM) counterparts.
Emulators are usually validated at the summary statistics level. In this study, we assess the emulator performance by performing full-field inference.  
We first show that analyzing PM convergence maps using the LPT forward model results in biased posterior estimates, highlighting simulation mismatch. 
Then, we demonstrate that the transformed simulations can recover the true posterior.

\section{Conditional Optimal Transport Flow Matching}

\subsection{Flow Matching}
\label{sec:FM}
FM \cite{lipman2023flowmatchinggenerativemodeling, albergo2023stochasticinterpolantsunifyingframework} is a method for training Continuous Normalizing Flows (CNFs) \citep{chen2019neuralordinarydifferentialequations}. CNF aims to continuously transport a source distribution $p_0$ into a target distribution $p_1$ over time through a diffeomorphic map $\phi:[0,1] \times \mathbb{R}^d \to \mathbb{R}^d$. This map is defined as the solution to the Ordinary Differential Equation (ODE):
\begin{align}
    dx = v_t(x)dt, \: \: \: \phi_0(x) = x_0,
\end{align}
where $x_0 \sim p_0$ are samples from the source distribution, and $v: [0,1] \times \mathbb{R}^d \to \mathbb{R}^d$ is the time-dependent velocity field guiding the transport.
Intuitively, the velocity field tells each particle $x$ in the distribution how to move at every time $t$ to gradually reshape $p_0$ into $p_1$. The intermediate distribution at any time $t$, denoted $p: [0,1] \times \mathbb{R}^d \to \mathbb{R}$, and known as the marginal probability path, is the pushforward
\footnote{The pushforward operator \# is defined by the change of variables formula $[\phi_t]_\# (p_0) = p_0(\phi_t^{-1}(x)) \det \left[ \partial_x \phi_t^{-1}(x) \right]$.}
of $p_0$ by the map $\phi_t$: $p_t = [\phi_t]_\# (p_0)$.
As long as the velocity field satisfies the continuity equation, the evolving distribution is a valid probability density function.

Once the velocity field $v_t$ is learned, generating samples from the target distribution simply requires integrating the ODE forward in time starting from samples of $p_0$. Conversely, one can reverse the process by integrating backward in time. If the density $p_1$ is required, it can be recovered via $p_1 = [\phi_1]_\# (p_0)$. 

In practice, FM turns learning $v_t$ into the following regression task
\begin{align}
    \mathcal{L}_{\text{FM}}  = \mathbb{E}_{t\sim\mathcal{U}[0,1], p_t(x)}\parallel v_t(x) - v_\varphi(t,x) \parallel^2,
\end{align}
where $v_\varphi$ is a Neural Network (NN). 
Because neither $p_t$ nor $v_t$ is known a priori, \citealp{lipman2023flowmatchinggenerativemodeling} proposed an alternative \textit{Conditional Flow Matching} (CFM) loss which shares the same gradients:
\begin{align}
    \mathcal{L}_{\text{CFM}}  = \mathbb{E}_{t\sim\mathcal{U}[0,1], q(z), p_t(x|z)}\parallel v_t(x|z) - v_\varphi(t,x) \parallel^2,
    \label{eq:cfm}
\end{align}
with $z$ a conditioning variable, $p_t(x|z)$ the conditional probability path such that $p_t(x) = \int p_t(x|z)q(z) dz$, and $v_t(x|z)$  the corresponding conditional velocity field.
As explain in \citealp{tong2024improvinggeneralizingflowbasedgenerative}, to bridge two arbitrary distributions, one can define $q(z)$ to generate pairs of samples $(x_0,x_1)$, for instance $q(z) = p_0(x_0)p_1(x_1)$. Then the conditional distribution $p_t(x|z)$ can be taken as a simple interpolating distribution, such as a Gaussian with mean along the straight path between $x_0$ and $x_1$: $\mu_t(z) = (1-t)x_0 + t x_1$, with a small variance \( \sigma^2 \). Specifically, as $\sigma \to 0$, $p_t$ at $t=0$ and  $t=1$ approaches the true source distribution $p_0$ and the true target distribution $p_1$, respectively.

Note that multiple velocity fields can transform 
$p_0$ into $p_1$, and enforcing only straight-line conditional probability paths between unpaired samples does not ensure that the velocity field minimizes the transport cost.  Additionally, FM only transports marginal to marginal distributions.

\begin{figure*}[t]
    \centering
    \includegraphics[width=\textwidth]{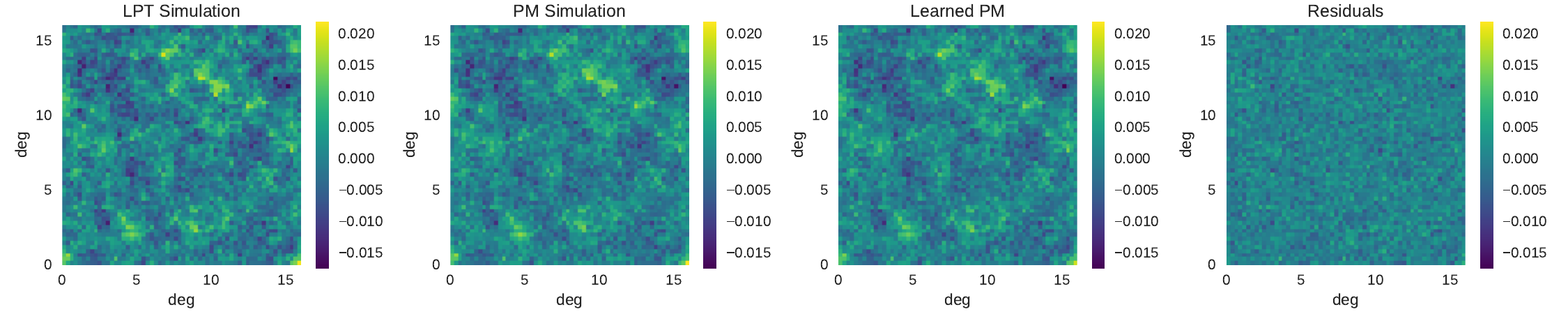}
    \caption{From left to right: An example of LPT simulations with noise; The corresponding PM simulation sharing the same noise, cosmological parameters and initial conditions as the LPT simulation; The LPT simulations optimally transporting onto the PM space; the residuals, i.e., the difference of the learned PM map with the true PM map without noise.
    }
     \label{fig:convergence_maps}
\end{figure*}
\subsection{Conditional Optimal Transport}
\label{sec:cot}
FM enables us to learn a deterministic transport map $\phi_1$ such that a new sample $x_1$ can be obtained from $x_0$ via $x_1 = \phi_1(x_0)$.  In this work, however, we seek the transport map that minimizes a specific cost, ensuring that the transformation from $x_0$ to $x_1$ retains essential information without unnecessary alterations. 
Moreover, this transformation must preserve the conditioning: each generated PM map should share the cosmological parameters of its corresponding LPT input.
Conditional Optimal Transport (COT) offers a rigorous framework for this goal by extending classical OT to the setting where one seeks to bridge conditional distributions across different parameters in an amortized way \cite{hosseini2024conditionaloptimaltransportfunction, kerrigan2024dynamicconditionaloptimaltransport}.

Formally, let $\Theta$ denotes the parameter space (e.g., cosmological parameters) and $X$ the observation space (e.g., convergence maps). We denote by $p_0(\theta,x)$ with $\theta \in \Theta$ and $x \in X$, the joint source distribution (LPT forward model), and by 
$p_1(\theta,x)$ that of the target (PM forward model), with both sharing the same prior $p(\theta)$ over parameters. To preserve conditioning during transport, we define a triangular map $\phi: \Theta \times X \rightarrow \Theta \times X$ of the form 
\begin{align}
    \phi(\theta, x) = \left(\theta, \phi_X(\theta, x)\right),
\end{align}
where $\phi_X: \Theta\times X \to X$.
A key result from \citealp{baptista2023conditionalsamplingmonotonegans}, shows that if $p_1 = [\phi]_\#(p_0)$, then for almost every $\theta$,  $\left[\phi_X(\theta, \cdot)\right]_\# \left(p_0(\cdot|\theta)\right) = p_1(\cdot|\theta)$.
That is, a single triangular map transports all likelihoods.

COT seeks such a triangular map $\phi$ that minimizes a displacement cost. In our case, we consider the cost  $c(x_0, \theta_0, x_1, \theta_1) = \|x_0 - \phi_X(\theta_0, x_0)\|^2$,
leading to the following conditional Monge problem:
\begin{align}
    \underset{\phi_X}{\inf} \int \|x - \phi_X(\theta, x)\|^2 p_0(\theta, x)dx d\theta,
\end{align}
subject to the constraint that $p_1 = [\phi]_\#(p_0)$.
If there exists a unique transport map $\phi^*$ that solves this problem, then $p_t = [\phi_t^*]_\# (p_0)$, with $\phi_t^* = (1-t)\text{Id} +t\phi^*$, is the conditional McCann interpolation \cite{kerrigan2024dynamicconditionaloptimaltransport}. This probability path and the corresponding triangular velocity field
$v_t(\phi^*_t(\theta, x)) = (0, \phi^*_X(\theta, x) - x)$, solve the COT problem. This result, known as the conditional Benamou-Brenier theorem, establishes the equivalence between the static and dynamic COT problem \cite{kerrigan2024dynamicconditionaloptimaltransport}.

Hence, to learn the velocity that solves the COT problem, we use the loss introduced in \autoref{eq:cfm} with the same straight-line conditional probability paths and the same corresponding conditional velocity field $v_t(x|z) = x_1 - x_0$. However, to ensure that the resulting velocity field corresponds to the solution of the COT problem, the sample pairs must be built as $x_1 = \phi^*_X(\theta_0, x_0)$ with $\phi^*_X$ the optimal conditional Monge map. Moreover, to preserve the conditioning during transport, the learned velocity field must be triangular $v_t(x, \theta) = (0, v_\varphi(t, x, \theta))$, with $ v_\varphi(t, x, \theta)$ the NN, so that $\theta$ remains unchanged. 

In practice, solving the static COT problem at the dataset scale is computationally intractable. We therefore use a minibatch approximation, solving COT separately on each minibatch \citep{fatras2021minibatchoptimaltransportdistances, tong2024improvinggeneralizingflowbasedgenerative}. Finally, to further ease optimization under finite sample constraints, we use a relaxed cost of the form $c(x_0, \theta_0,x_1, \theta_1) = |\theta_0 - \theta_1| + \epsilon |x_0 - x_1|$. As $\epsilon \to 0$, \citealp{hosseini2024conditionaloptimaltransportfunction} show that we recover the triangular map solving the conditional Monge problem. 
Note that when paired simulations from both models are available, finding $\phi^*$ becomes unnecessary. 
\begin{figure}[h]
    \centering
    \includegraphics[width=0.75\columnwidth]{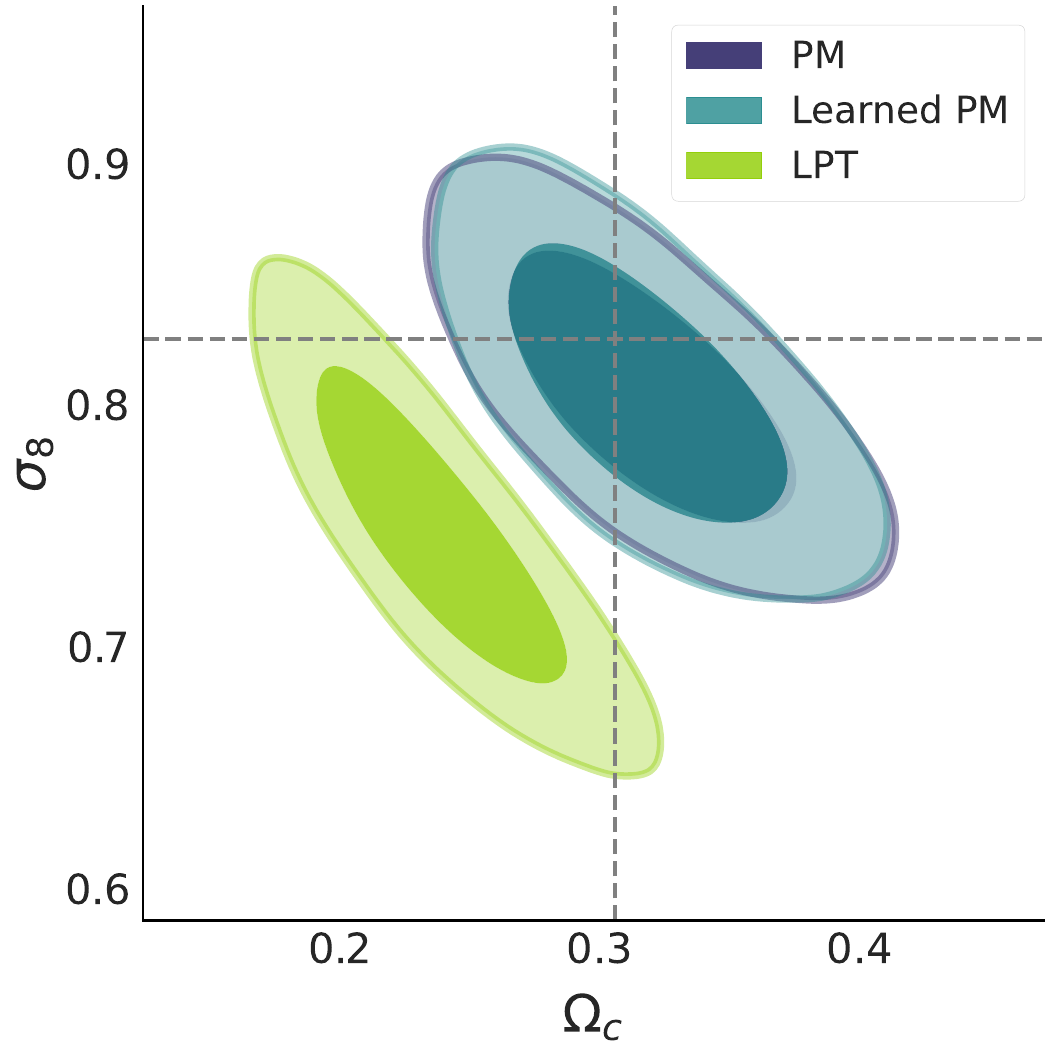}
    \caption{Posterior distributions of the cosmological parameters evaluated on the PM map shown in \autoref{fig:convergence_maps}, and learned from  LPT simulations (green), PM simulations (purple), and emulated simulations (blue).}
    \label{fig:contour_plot}
\end{figure}
\begin{figure}
    \centering
    \includegraphics[width=0.75\columnwidth]{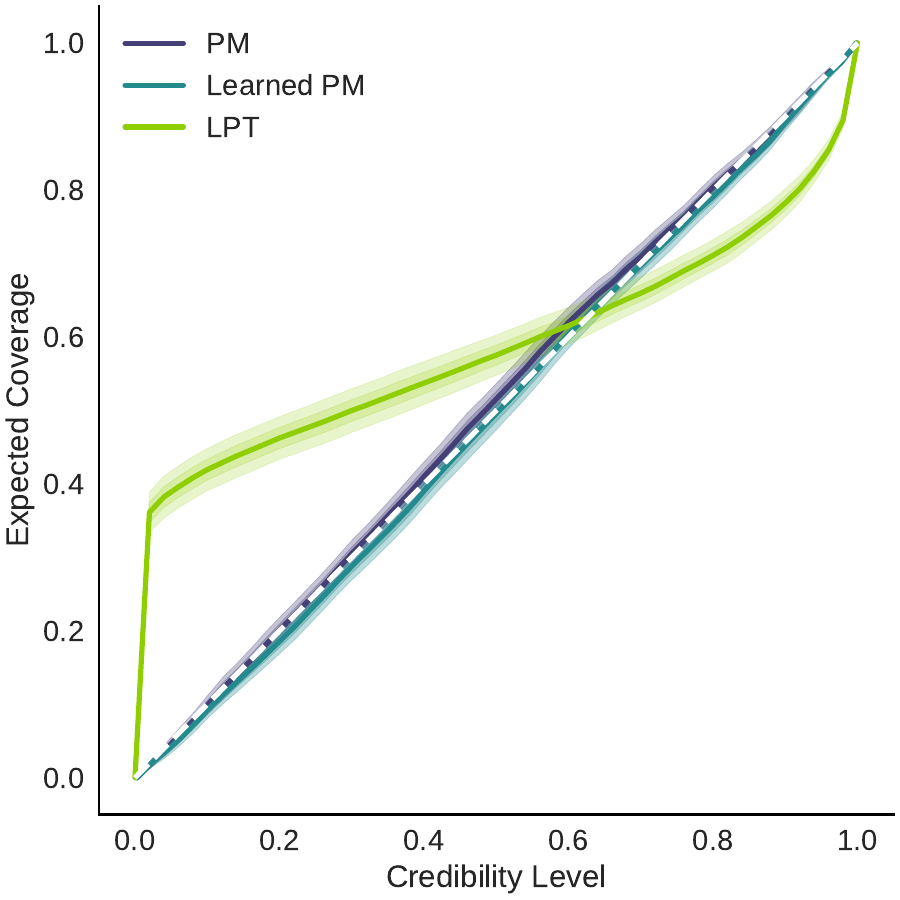}
    \caption{TARP expected coverage test over $500$ simulations, with error bars from $100$ bootstrap iterations. The dashed white line shows ideal coverage. Posteriors from PM maps analyzed with the correct model (purple) closely match this curve, while using the LPT model (green) leads to bias. COT-FM restores accurate coverage (blue), demonstrating effective emulation of PM maps.
}
     \label{fig:tarp}
\end{figure}
\section{Experiments}
\subsection{Simulation Models}
We generate two suites of simulations: second‐order LPT and PM N‐body, using \url{JaxPM}\footnote{https://github.com/DifferentiableUniverseInitiative/JaxPM}.  Both sets of simulations are run in a cubic box of comoving size $500\times500\times5000\ \mathrm{Mpc}\:h^{-1}$ discretized into a $60\times60\times200$ voxel grid, corresponding to a transverse resolution of $8.3 \mathrm{Mpc}\:h^{-1}$ and
a radial resolution of $25 \mathrm{Mpc}\:h^{-1}$. 
Displacements of the PM simulations are first computed using first-order LPT up to scale factor $0.1$, and then according to the Poisson equation. The particle displacement is interpolated onto a 3D density grid using the Cloud-in-Cell (CIC) scheme, and we treat the resulting volume as our light cone. 
Convergence maps are then computed under the Born approximation by integrating density fluctuations along the line of sight. The resulting maps span an area of $16^\circ \times 16^\circ$ on the sky, discretized into $60 \times 60$ pixels (pixel scale $\approx 0.27^\circ$).
We assume a single source redshift distribution given by $n(z)\propto z^{0.5}e^{-(z/2.0)^{1.0}}$ with a galaxy density of $ n_{\rm gal} = 30\ \mathrm{arcmin}^{-2}$.  Shape noise is added per pixel using
$\sigma_{\rm pix}=\sigma_e/\sqrt{n_{\rm gal}\,A_{\text{pix}}}$,
where $\sigma_e=0.26$ and $A_{\text{pix}}= 256\ \mathrm{arcmin}^{2}$. Cosmological parameters $\theta=(\Omega_c,\sigma_8)$ are drawn from independent Gaussians, $\Omega_c\sim\mathcal{N}(0.3,0.05^2)$ and $\sigma_8\sim\mathcal{N}(0.8,0.05^2)$. An example of the different convergence maps is shown in \autoref{fig:convergence_maps}. 

Although we acknowledge that this is a simplistic setup, it is sufficient for initial testing purposes. A more complete study, with more realistic simulations will be presented in a forthcoming paper.

\subsection{Implicit full-field inference}
We perform full-field inference using an implicit inference approach that allows for amortized inference across various observations, which is useful for assessing the robustness of our correction method. However, sampling schemes such as Hamiltonian Monte Carlo \cite{neal2011mcmc, betancourt2017conceptual} can also be applied to our corrected forward model.

Convergence maps are first compressed into a two-dimensional summary statistic by training a Convolutional Neural Network (CNN) under the Variational Mutual Information Maximization (VMIM) loss \citep{Jeffrey_2020}, which has been shown to build near-sufficient statistics for parameter inference \citep{lanzieri2025optimalneuralsummarisationfullfield}.
Then, we perform Neural Posterior Estimation (NPE) \citep{papamakarios2018fastepsilonfreeinferencesimulation} by training a Normalizing Flow (NF) on the compressed summaries to learn the posterior distribution over cosmological parameters.
All training details, including the number of simulations, are provided in \autoref{sec:training_details}.

\subsection{Results}
We begin by highlighting the simulation mismatch at the pixel level. To this end, we apply our implicit inference framework: For the LPT simulations, we train the CNN compressor and NF posterior estimator using only LPT samples. For the PM simulations, we repeat the same procedure but using PM samples.
Once trained, we evaluate the two inferred posterior distributions on a fixed PM fiducial convergence map (shown in \autoref{fig:convergence_maps}). 
The results in \autoref{fig:contour_plot} illustrate a significant bias when analyzing PM data with a model trained on LPT simulations.

To correct this bias, we apply COT-FM.
As explained in \autoref{sec:cot}, COT-FM learns a COT map that transforms LPT simulations into PM-like simulations, preserving the associated cosmological parameters and finding the minimal correction to the LPT simulation needed to match its PM counterpart, a process that is effective even when only two unpaired sets of simulations are available.
Our mapping is parameterized using a $2$D conditional U-Net \cite{ronneberger2015unetconvolutionalnetworksbiomedical} trained on two noisy datasets: the set of LPT simulations and the set of PM simulations. At each training step, we solve the COT problem on the mini-batch to effectively pair each LPT and PM simulation, allowing us to learn the velocity field that solves the dynamic COT problem.
Once the velocity field is learned, we integrate the forward ODE to transport LPT samples into their corrected PM-like counterparts.
An example of a transported convergence map is shown in \autoref{fig:convergence_maps}.

As a first validation, we project the original LPT, learned PM, and true PM simulations into summary space using the PM-trained compressor. As shown in \autoref{fig:summary_space}, the learned PM simulations closely align with the true PM simulations, while differing from the original LPT simulations.
We complete the evaluation by performing implicit full-field inference, i.e., training the CNN compressor to build sufficient statistics and training the NF posterior estimator using the learned PM convergence maps. We evaluate the posterior distribution on the same fixed convergence maps as for LPT and PM posteriors. The results in \autoref{fig:contour_plot} show that COT-FM effectively corrects the inference bias, demonstrating the emulator accuracy. To further validate the posterior distributions across multiple observations, we use the TARP \cite{lemos2023samplingbasedaccuracytestingposterior} Expected Coverage Probability (ECP) test. The ECP corresponds to the expected number of posterior samples that fall in a credible region of credibility level $1-\alpha$. TARP provides a method to estimate this ECP such that it is precisely equal $1-\alpha$ for all $\alpha \in [0,1]$ if and only if the posterior distribution is calibrated. In \autoref{fig:tarp}, we apply TARP to the posterior distributions obtained from the three simulation models and evaluated on $500$ PM observations. We show that, unlike the LPT maps, the learn maps recover the true posterior distributions, as indicated by the ECP falling along the diagonal.
Finally, additional posterior distributions are displayed in \autoref{fig:contour_plots_appendix} with their corresponding convergence maps displayed in \autoref{fig:convergence_maps_appendix}.

\section{Conclusion and Discussion}
We have presented a new method to accurately emulate simulations at the pixel level. Instead of modeling complex simulation dynamics end-to-end, our approach focuses on transforming simulation likelihoods. This significantly reduces the complexity of the learning task, requiring fewer training simulations and lowering the risk of generating unphysical outputs. In addition, by framing the problem as a transport between distributions under the COT constraint, the method applies minimal yet targeted corrections to align forward models. We also note that because the transport operates on likelihood distributions, the emulator is naturally conditioned on cosmological parameters. The learned transformation is differentiable, making it suitable for explicit full-field inference that relies on gradient sampling schemes. Finally, and maybe most importantly, the COT-FM framework allows training on unpaired datasets, which increases flexibility when exact simulation pairs are not available.
Applied to a toy weak lensing example, we demonstrated the effectiveness of this method by successfully recovering the full-field posterior distribution using our emulated convergence maps.

In future work, we plan to more rigorously investigate the limitations of this approach, for instance, when the LF and HF distributions differ significantly or in high-noise regimes. We also aim to extend our experiments to more realistic convergence maps from Stage-IV surveys, for example, by mapping PM simulations to hydrodynamic simulations, incorporating baryonic effects, and modeling observational systematics. Additionally, we will explore transporting full simulation volumes directly. 
We note that, for this method to be effective, it is crucial that COT-FM training works well, even with a limited number of high-quality simulations. An ablation study on the size of the training set will be conducted in future work.
Another potential direction is to condition the transport not only on cosmological parameters $\theta$, but also on the initial conditions. This would ensure that paired simulations share the same IC, potentially yielding even more minimal transports. 
We believe that this framework also offers a promising avenue for future extensions, in which it could be used as a correction tool for bridging simulations to real data. However, it will require rethinking assumptions made in this study, since our Universe has a fixed but unknown set of cosmological parameters. 

\section*{Acknowledgements}
The authors would like to thank Nikolay Malkin and Alexandre Adam for their valuable discussions. We also thank the reviewers of the ML4Astro 2025 workshop for their constructive feedback.
This work is partially supported by Schmidt Futures, a philanthropic initiative founded by Eric and Wendy Schmidt as part of the Virtual Institute for Astrophysics (VIA). The work is in part supported by computational resources provided by Calcul Quebec and the Digital Research Alliance of Canada. Y.H. and L.P. acknowledge support from the Canada Research Chairs Program, the National Sciences and Engineering Council of Canada through grants RGPIN-2020-05073 and 05102.

\bibliography{example_paper}
\bibliographystyle{icml2025}

\clearpage

\appendix
\section{Training details and model architectures}
\label{sec:training_details}
\paragraph{Implicit full-field inference} is performed by first compressing $60 \times 60$ pixels convergence maps into $2$-dimensional summary statistics using a $2$D Convolutional Neural Network (CNN) composed of three convolutional layers. The CNN is trained using the VMIM loss function \cite{Jeffrey_2020} on a dataset of $60\,000$ simulations. To compute the VMIM loss function, we use a RealNVP NF \cite{dinh2017densityestimationusingreal} with $4$ coupling layers.
After training, the compressor is used to compress a new set of $40\,000$ simulations. A new RealNVP with $4$ coupling layers is then trained on the compressed summary statistics using the NPE loss function to approximate the posterior distribution.
The architectures of both the compressor and the NFs are kept fixed across all experiments.

\paragraph{COT-FM} is performed using a 2D conditional U-Net \cite{ronneberger2015unetconvolutionalnetworksbiomedical} from the \texttt{Diffusers} library\footnote{\url{https://huggingface.co/docs/diffusers/}}, which we adapt to put cosmological parameters as an additional conditioning input alongside the time. The U-Net is trained under the COT-FM loss to learn the marginal velocity field, using a dataset of $90\,000$ simulations (for both LPT and PM), different from the datasets used in implicit full-field inference.
We train with a batch size of $200$ simulations and use the exact linear programming Earth Mover's Distance (EMD) algorithm from the \texttt{POT} library \cite{flamary2021pot} to solve the COT problem in each batch. For the transport cost, we set $\epsilon = 0.1$.  Note that larger batch sizes and smaller $\epsilon$ improve the OT approximation. Although we did not fine-tune these parameters in this study, we plan to address them in future work. Finally, we integrate the learned velocity field using the Dopri5 ODE solver \cite{dormand1980family, shampine1986some} from the \texttt{diffrax} library \cite{kidger2021on}.

We trained COT-FM for seven hours on an NVidia A100SXM4.

\newpage

\section{Additional figures}
\label{sec:additional_fig}

\begin{figure}[!h]
    \centering
    \includegraphics[width=0.75\columnwidth]{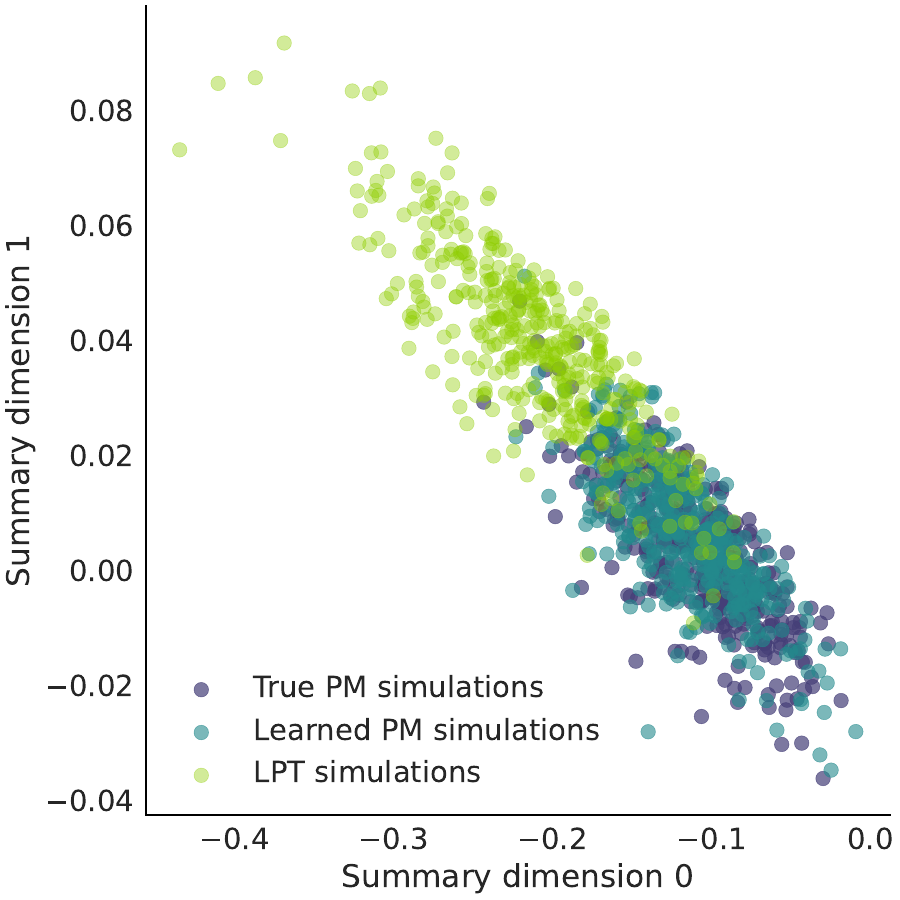}
    \caption{Simulations comparison in summary space. Using the compressor trained on PM maps, we compress PM maps (purple), learned PM maps (blue), and LPT maps (green). We show that our emulated PM maps align with the true PM ones.
    }
     \label{fig:summary_space}
\end{figure}

\begin{figure*}
  \centering
  \begin{minipage}{0.88\columnwidth}
    \centering
    \includegraphics[width=0.88\columnwidth]{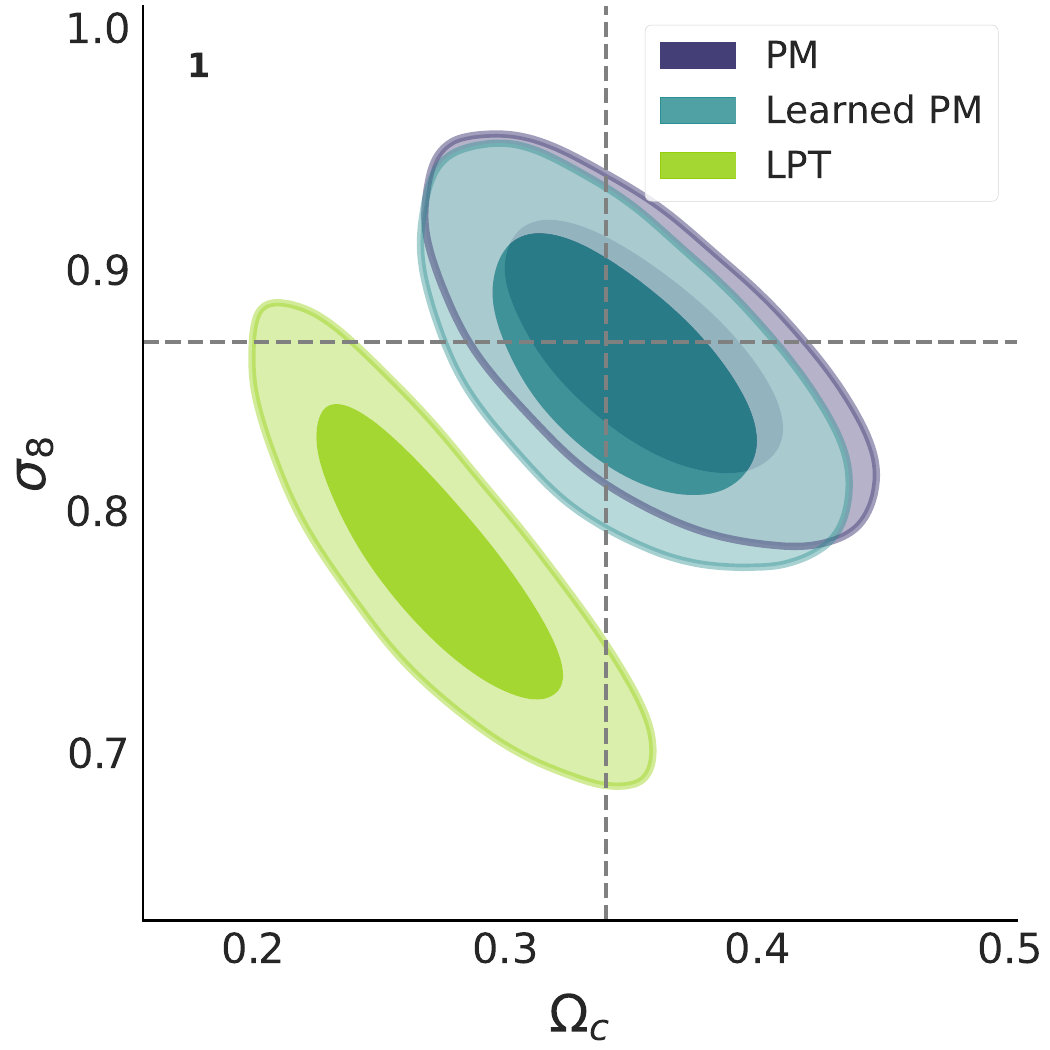}
  \end{minipage}
  \hfill
  \begin{minipage}{0.88\columnwidth}
    \centering
    \includegraphics[width=0.88\columnwidth]{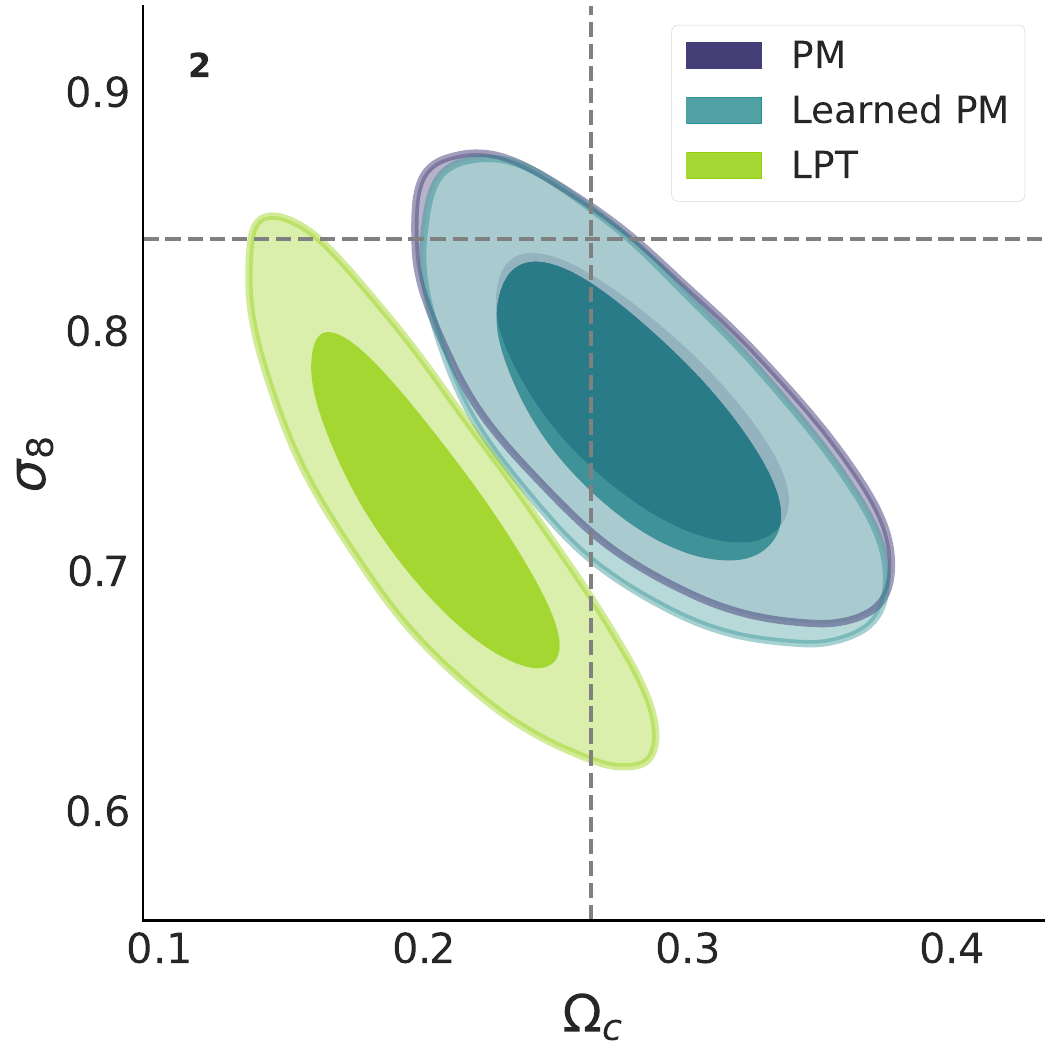}
  \end{minipage}
  
  \vspace{0.5cm}
  
  \begin{minipage}{0.88\columnwidth}
    \centering
    \includegraphics[width=0.88\columnwidth]{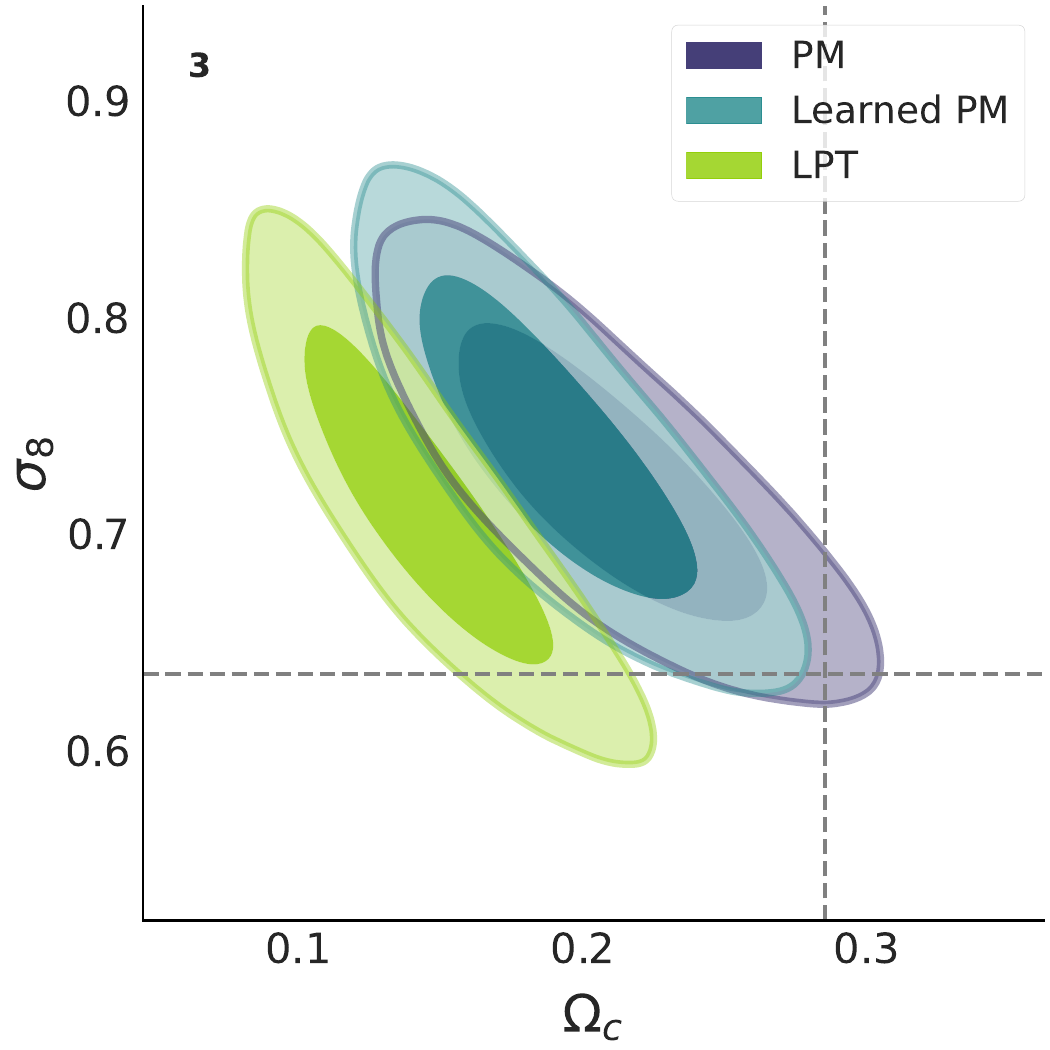}
  \end{minipage}
  \hfill
  \begin{minipage}{0.88\columnwidth}
    \centering
    \includegraphics[width=0.88\columnwidth]{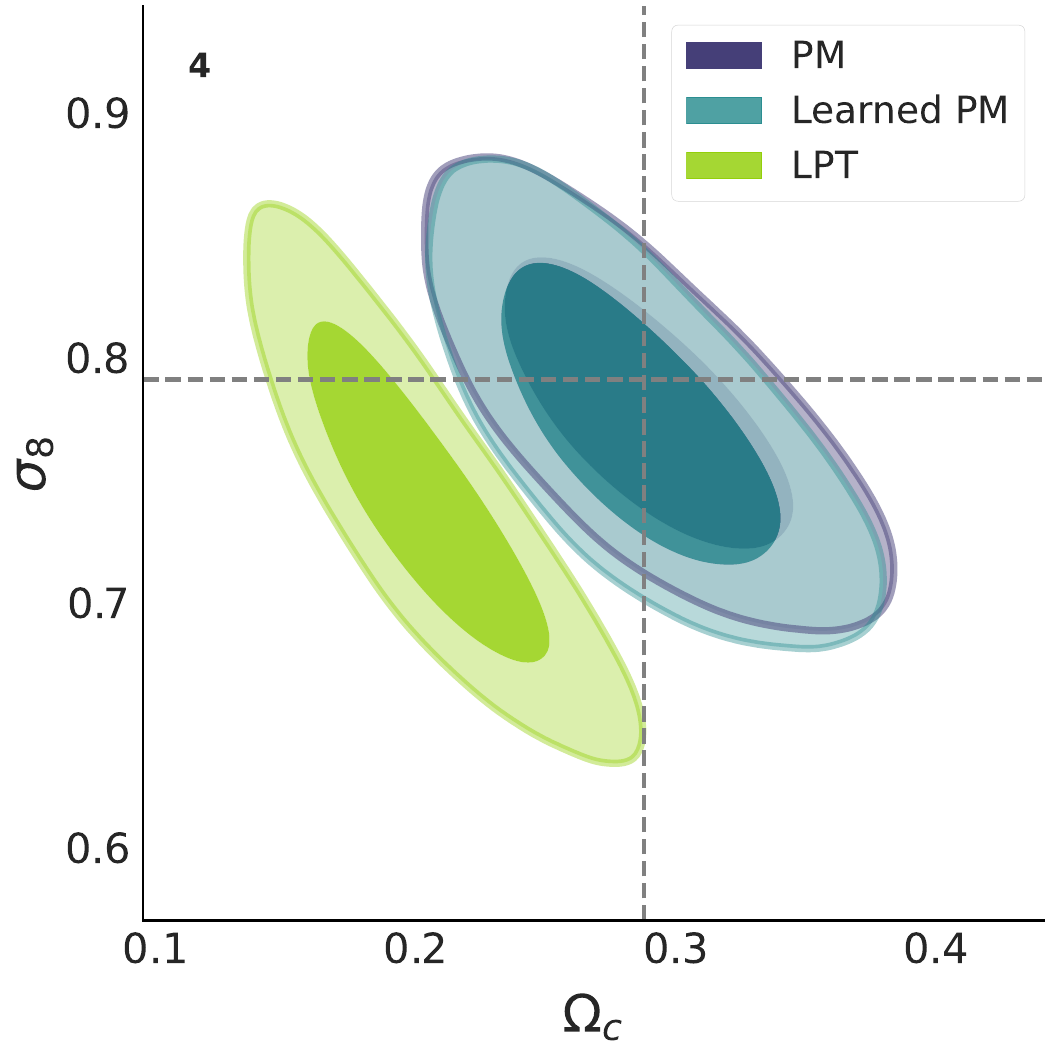}
  \end{minipage}
  
  \vspace{0.5cm}
  
  \begin{minipage}{0.88\columnwidth}
    \centering
    \includegraphics[width=0.88\columnwidth]{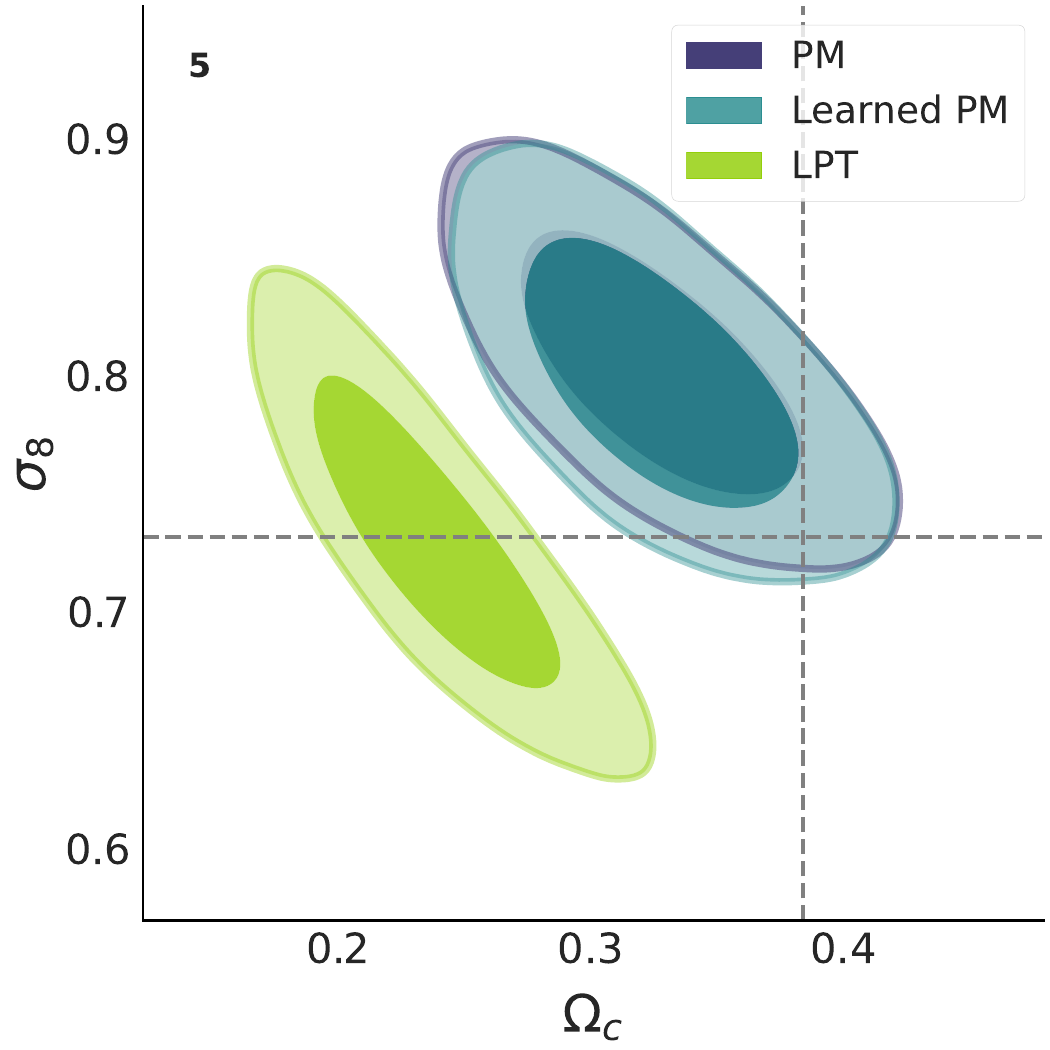}
  \end{minipage}
  \hfill
  \begin{minipage}{0.88\columnwidth}
    \centering
    \includegraphics[width=0.88\columnwidth]{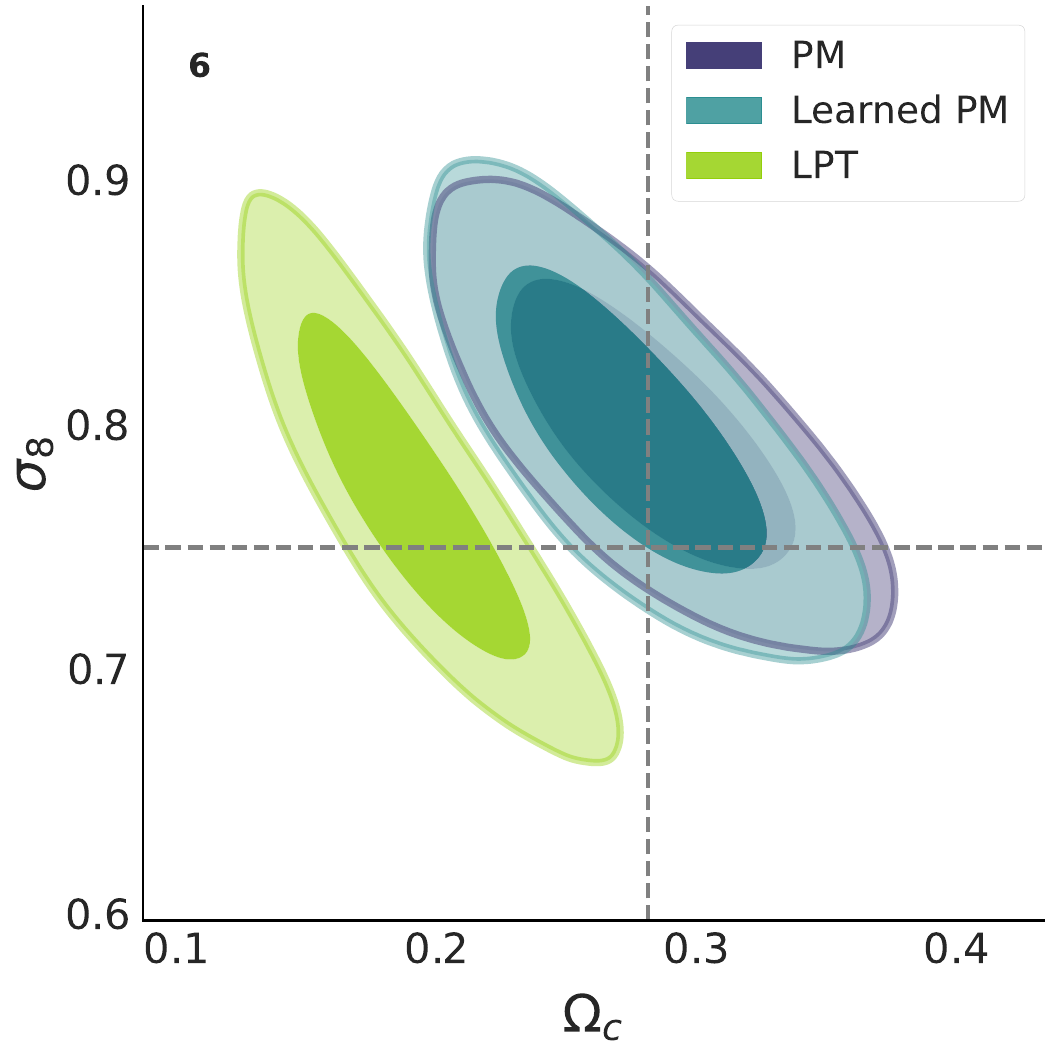}
  \vspace{0.5cm}
  \end{minipage}

  \caption{Additional posterior distributions of the cosmological parameters evaluated on the PM fiducial maps shown in \autoref{fig:convergence_maps_appendix}. The green contour corresponds to the posterior distribution learned using LPT simulations. The purple one corresponds to the posterior distribution learned using PM simulations. The blue one is the posterior distribution learned from corrected LPT simulations, i.e., PM-like simulations. This highlights the impact of model misspecification on the cosmological constraints and how COT-FM solves this problem.}
  \label{fig:contour_plots_appendix}
\end{figure*}

\begin{figure*}  
    \centering
    \begin{subfigure}{\textwidth}
        \centering
        \includegraphics[width=\textwidth]{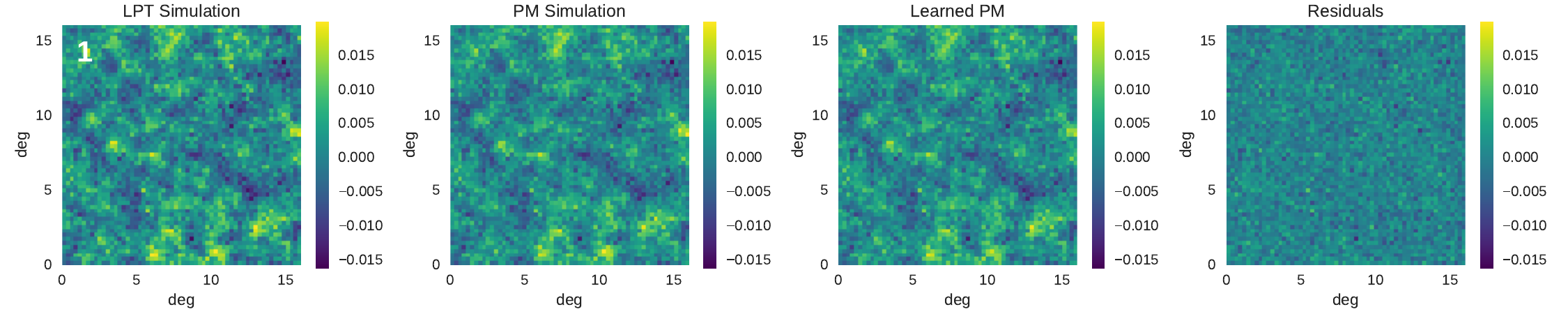}
    \end{subfigure}

    \begin{subfigure}{\textwidth}
        \centering
        \includegraphics[width=\textwidth]{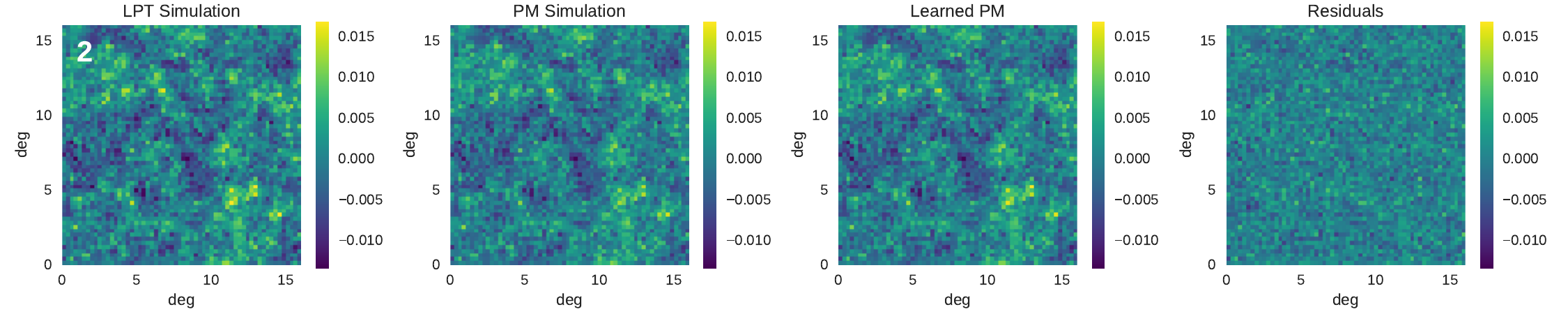}
    \end{subfigure}

    \begin{subfigure}{\textwidth}
        \centering
        \includegraphics[width=\textwidth]{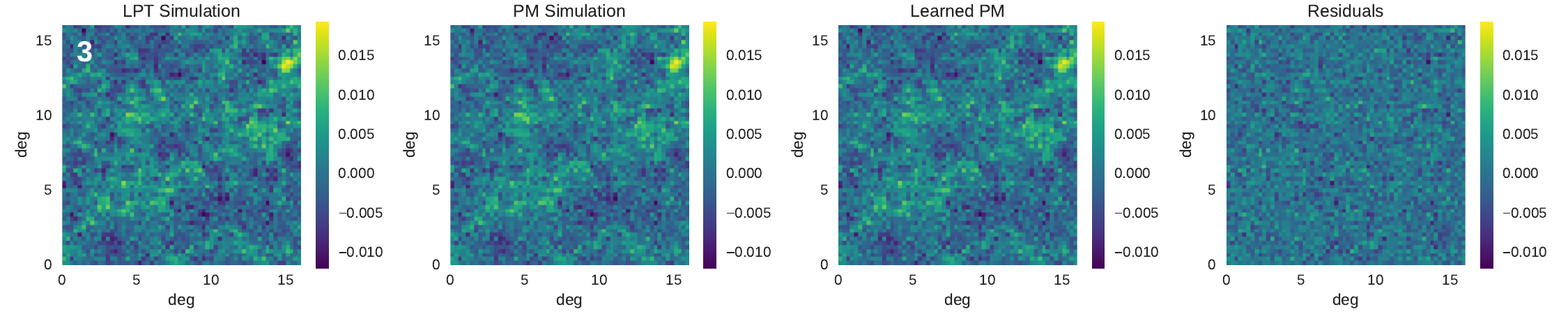}
    \end{subfigure}

    \begin{subfigure}{\textwidth}
        \centering
        \includegraphics[width=\textwidth]{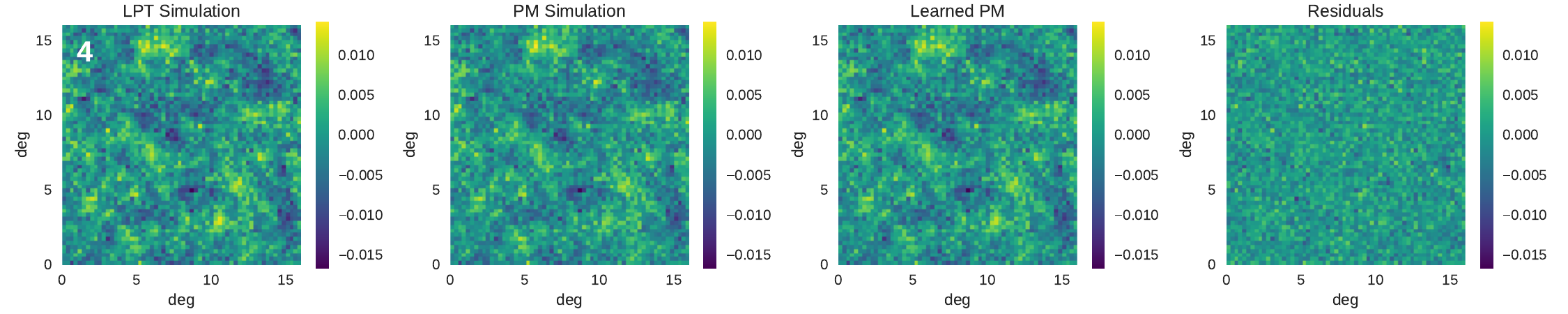}
    \end{subfigure}

    \begin{subfigure}{\textwidth}
        \centering
        \includegraphics[width=\textwidth]{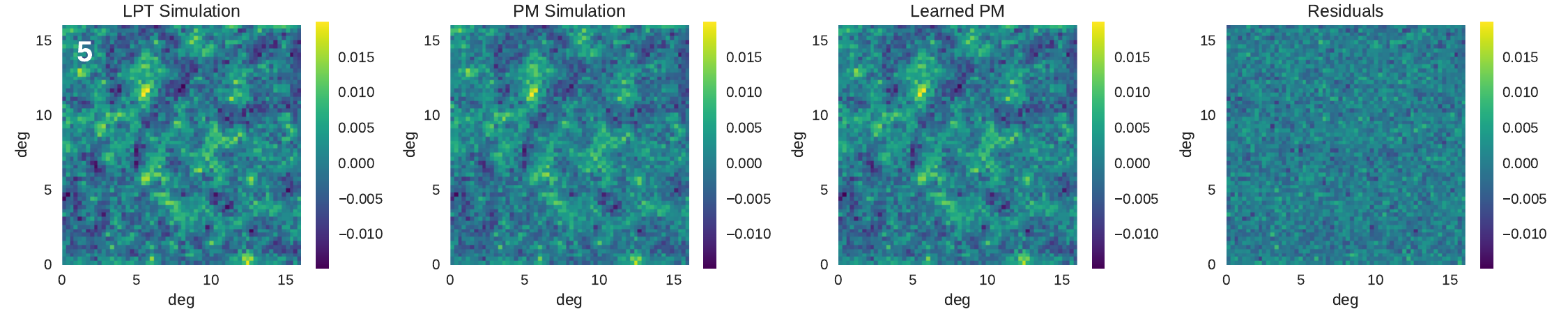}
    \end{subfigure}

    \begin{subfigure}{\textwidth}
        \centering
        \includegraphics[width=\textwidth]{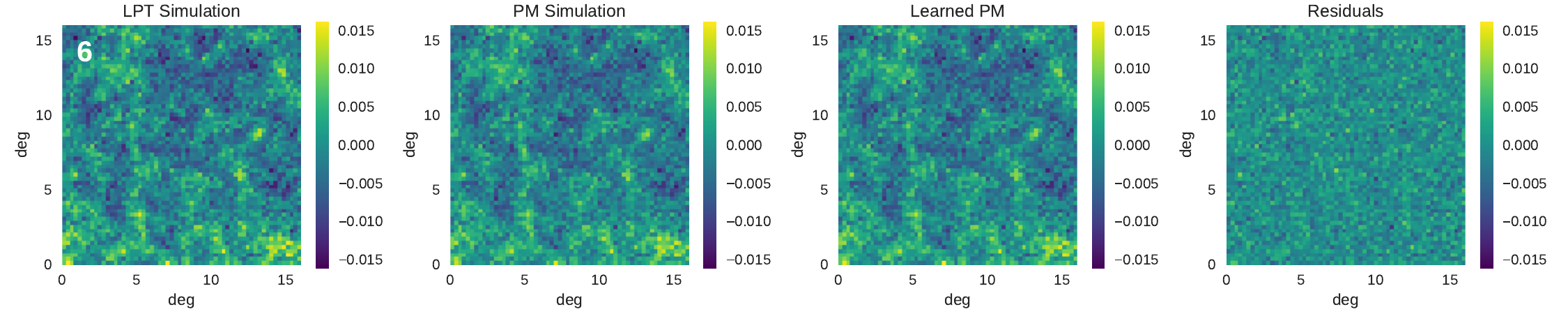}
    \end{subfigure}

    \caption{Additional convergence maps. From left to right: LPT simulations with noise; The corresponding PM simulation sharing the same noise, cosmological parameters and initial conditions as the LPT simulation; The LPT simulations optimally transporting onto the PM space; the residuals, i.e., the difference of the learned PM map with the true PM map without noise. }
    \label{fig:convergence_maps_appendix}
\end{figure*}

\end{document}